\documentclass[longbibliography,reprint,amsmath,amssymb,aps,nofootinbib]{revtex4-2}

\usepackage{graphicx}
\usepackage{natbib}

\begin{document}

\title{Universality of satellites in the breakup of a stretched fluid bridge}
\author{Anna Frishman$^{1,2}$}
\author{Daniel Lecoanet$^{3,4,2}$} 
\affiliation{$^1$Technion Israel Institute of Technology, 32000 Haifa, Israel}
\affiliation{$^2$Princeton Center for Theoretical Science, Princeton University, Princeton, New Jersey 08544, USA}
\affiliation{$^3$Department of Engineering Sciences and Applied Mathematics, Northwestern University, Evanston 60208, IL, USA}
\affiliation{$^4$CIERA, Northwestern University, Evanston 60201, IL, USA}

\begin{abstract}
As a fluid object breaks, it often leaves behind satellite fragments. Here we show that satellite formation can follow universal dynamics, leading to robust satellite sizes.
Specifically, we consider the breakup of a slowly stretched fluid bridge, which we realize experimentally using a soap-film bubble suspended between two plates. Combining experiments and one-dimensional simulations, we show that a main satellite bubble always forms as the bridge breaks. We discover that the size of the bubble is highly reproducible and can be dramatically increased by stretching the bridge faster or increasing its volume. The satellite size is a simple function of two non-dimensional parameters: the normalized volume of the bridge and the Weber number, measuring inertia due to stretching as compared to surface tension. These observations can be explained by tracing the bridge evolution over a series of dynamical stages in which the bridge: (i) closely follows a sequence of equilibrium bridge configurations; (ii) stretches as it begins to breakup after reaching an unstable equilibrium; and (iii) follows a universal breakup solution. The last stage takes place over a finite region, the corresponding length scale determined by stretching during the previous stage. This length scale controls the satellite size, and the universality of the dynamics makes the system highly reproducible. This work suggests universal satellite formation dynamics may provide a route for understanding satellite bubble sizes in turbulent flows.
\end{abstract}

\maketitle

\section{Introduction}

The fragmentation of fluid objects is central to many industrial and natural processes, from ocean-atmosphere exchange in breaking waves \cite{veron_ocean_2015}, to the preparation of emulsions \cite{villermaux_fragmentation_2007}. Small satellite drops are often generated in the process, either in individual breakup events or through their sequence. 
Predicting the size and number distribution of satellites in fragmentation is thus a fundamental problem of practical importance. For breakup where viscous effects dominate over inertia, the sizes of satellites based on individual breakup events have been studied in detail~\cite{tjahjadi_satellite_1992,stone_dynamics_1994}. Their formation can be controlled to a high degree by controlling the external flow~\cite{christopher_microfluidic_2007}. When viscous forces are comparable to inertia, it was recently discovered that satellite sizes can be universal~\cite{ganan_revision_2017}. However, when viscous forces are negligible compared to inertia (high Reynolds number flow)~\cite{mansour_satellite_1990,rodriguez_breakup_2006,kooij_what_2018}, initial conditions and external perturbations seem to be important. A well known example is the Rayleigh-Plateau instability, responsible for the breakup of a jet into drops. The size and number of fragments then depends on the initial perturbation, which determines the relevant unstable mode~\cite{mansour_satellite_1990,eggers_nonlinear_1997}. For high Reynolds number flows, especially for breakup driven by turbulence~\cite{vela_deformation_2021,riviere_sub_2021,riviere_capillary_2022,villermaux_fragmentation_2020}, such initial conditions are often unknown and control of the external flow can be difficult.  

While the size of satellites can be sensitive to the precise flow conditions, the dynamics around pinch-off points are often universal due to their singular nature~\cite{eggers_drop_2005,eggers_singularities_2015,deblais_viscous_2018}. Namely, the dynamics become independent of the initial conditions and spatial details away from the pinch-off point.\footnote{Exceptions to universal behaviour have also been observed~\cite{rallison_1978_numerical,doshi_persistence_2003}, e.g. due to asymmetries~\cite{turytsen_asymmetric_2009}.}
The universal pinch-off behavior occurs on small lengthscales, in many cases, much smaller than the size of satellites.
This suggests satellite sizes may be determined by processes unrelated to pinch-off.

A prototypical system that exhibits fragmentation is the fluid bridge, which has been studied since at least Plateau~\cite{plateau_experimental_1863}. Fluid bridges remain a subject of continued interest due to their ubiquity and practical and fundamental importance, see~\cite{montanero_review_2019} and references therein.
Refs.~\cite{cryer_collapse_1992,chen_dynamics_1997,robinson_observations_2001,goldstein_geometry_2021} studied fragmentation of soap film bridges between two open rings.
The bridge fragments, forming one main satellite bubble, when the separation between the rings is larger than a critical value.
While experiments and simulations show the size of the satellite bubble is reproducible, these works do not determine what sets this size.
For inertia-driven breakup, if the system is reflection-symmetric around the initial necking point, the breakup generically produces a satellite bubble instead of pinching off at a single point~\cite{zhang_nonlinear_1996,chen_dynamics_1997,gordillo_satellites_2007, huang_pinching_2019}.
The satellite forms because the inertia of the fluid builds up near the necking point, flattening the surface until it overturns, and sealing a satellite bubble upon breakup.

We extend this work to consider a fluid bridge between two solid plates, such that the volume of the bridge is conserved. It is then driven to breakup as it is slowly stretched by the movement of its end plates. This setup includes two new dimensionless control parameters---the non-dimensionalized volume and speed---and we determine the satellite size as a function of these parameters. We realize this set-up using a soap bubble placed between two solid plates, forming a soap-film bridge.
As the evolution is controlled by the dynamics of the air inside the bridge~\cite{robinson_observations_2001}, we accompany the experiments with simulations of a general fluid bridge, where only the air dynamics are modeled. These minimal simulations reproduce the main findings of our experiments, demonstrating the relevance to e.g. a liquid bridge. 
While previous works have studied fragmentation of liquid bridges and fixed-volume fluid bridges experimentally, e.g., \cite{kroger_stretching_1992, zhang_nonlinear_1996,zhuang_combined_2015}, and numerically, e.g., \cite{yildrim_deformation_2001,vincent_forced_2014}, none addressed satellite formation.

Each of our experiments is initiated with a cylindrical soap-film bridge, as shown in Fig.~\ref{fig:images}$(a)$. The initial aspect ratio is denoted by $L_0$. The plates are pulled apart as described in Appendix~\ref{sec:experiment}. At the plates, the bridge remains practically pinned to its initial height, denoted by $h_0$. We non-dimensionalize all length-scales using $h_0$. The initial volume of air inside the bridge, $V$, is given by $V/(h_0^3)=\pi L_0$ and it remains fixed throughout the evolution (Fig.~\ref{fig:images}$(b)$). Thus, $L_0$ is a non-dimensional control parameter for our experiments.

The breakup of the bridge is driven by surface tension. 
To quantify the relative importance of inertia and surface tension we introduce the Weber number $\text{We} = \rho v_p^2 h_0 /\sigma$, where $2v_p$ is a representative speed with which the plates separate, $\rho$ is the density of air inside the bridge and $\sigma$ is the surface tension. The We number is our second non-dimensional control parameter. 
Our experiments are performed in the regime of low Weber number, high Reynolds number,\footnote{Viscosity becomes important only at late stages of the breakup and around the pinch-off points, a known effect, e.g.~\cite{lister_capillary_1998,eggers_singularities_2015,Casterjon_plethora_2015}} low Ohnesorge number (small viscosity compared to surface tension effects) and low Bond number\footnote{The combination of low Weber number, high Reynolds number, and low Bond number is easily achieved for a soap film bridge. This regime was also achieved for liquid bridges in Ref.~\cite{kroger_stretching_1992}, which also formed satellite drops.} (see Appendix~\ref{sec:modeling}).

As a fixed-volume bridge is stretched, it eventually breaks, always leaving behind two spherical caps at the plates and one central satellite bubble,\footnote{Upon breakup the central bubble is attached to the two caps via fluid strings. Those strings then break into a multitude of very small satellite bubbles, which we do not consider here.} as in Fig.~\ref{fig:images}$(a)$ and Fig.~\ref{fig:images}$(b)$.  
Here, varying We and the non-dimesionalized volume, we reveal that the formation of the satellite
follows a universal route, resulting in robust satellite sizes, see Fig.~\ref{fig:images}. The corresponding universal dynamics are novel, occurring on a length scale which remains finite in time, see the third panel in Fig.~\ref{fig:images}$(b)$, and are unrelated to the dynamics at the pinch-off points. Moreover, we show that the volume of the resulting satellites has a simple functional dependence on the two control parameters and is an increasing function of both.   
This provides the first example in the high Reynolds number regime
where satellite sizes can be characterized without knowledge of the exact flow conditions and their control.  

\begin{figure*}[ht]
    \begin{centering}        \includegraphics{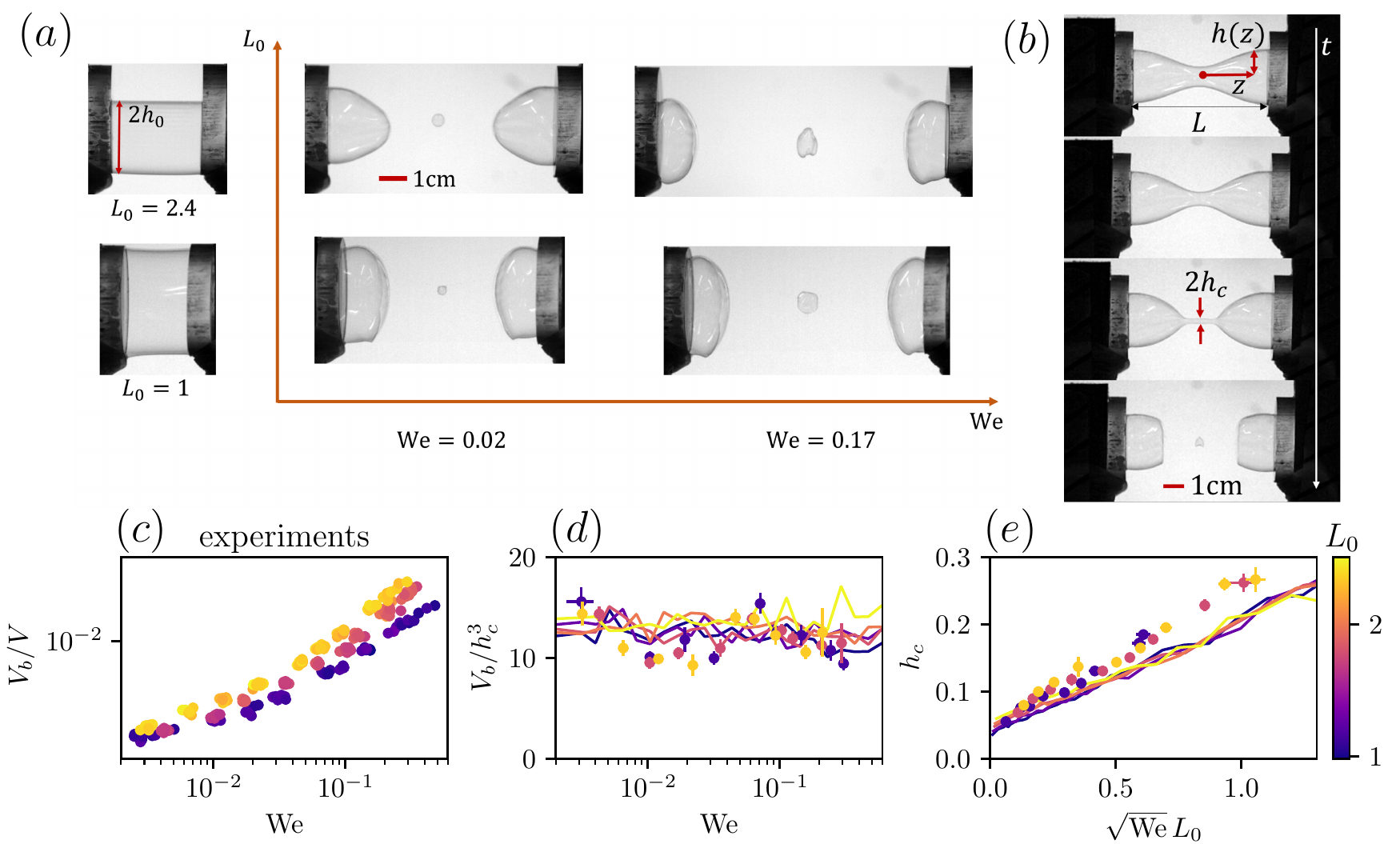}
        \caption{$(a)$ Soap-film bridge experiments. Left panels show initial conditions with different $L_0$, the ratio of the initial length to the radius $h_0$ of the cylindrical bridge. Right panels show the final state with a central satellite bubble for experiments with different $L_0$ and Weber number ${\rm We}$. We study bridges with typical radii $\sim 1\, {\rm cm}$. $(b)$ Snapshots of a bridge with $L_0=2.4$ pulled apart with ${\rm We}=0.02$ to form a central satellite bubble. The snapshots are evenly spaced in time. $(c)$ The volume of the satellite bubble normalized to the initial bubble volume as a function of ${\rm We}$; different colors show different $L_0$. Each dot represents a single experiment.
        $(d)$ Volume of the satellite bubble normalized by $h_c^3$, where $h_c$ is the radius of the center of the bridge when it is flat (see panel $(b)$). We find that $V_b\approx 13 h_c^3$ in both experiments and simulations.
        $(e)$ $h_c$ as a function of $\sqrt{{\rm We}}\, L_0$, in experiments and simulations. In all cases, we find $h_c$ is linearly proportional to $\sqrt{{\rm We}}\, L_0$ plus a constant term.
    In both plots, dots show the ensemble average of experimental results with error bars representing one standard deviation; lines show results from simulations.
        \label{fig:images}}
    \end{centering}
\end{figure*}

\section{Numerical Modelling} \label{sec:modelling}
We aim to capture the phenomenology of the breakup dynamics within the simplest possible model, focusing on satellite formation. This will allow us to probe the generality and robustness of the obtained results, and reveal their physical origin. In particular, we focus on the air dynamics within the bridge, and do not model the soap film itself. The bridge in our experiments remains axisymmetric for the bulk of the dynamics, as they are performed in the low Bond number regime. Its shape is thus captured by the radius along the shape, equivalent to the height $h(z)$ as measured from the axis of symmetry in the images, see Fig.~\ref{fig:images}$(b)$. To further simplify the model, we use the slender jet approximation~\cite{eggers_drop_1994}, which here can be justified both by the axisymmetry of the bridge, and by the fact that it becomes increasingly slender as the dynamics proceeds (Appendix~\ref{sec:modeling}). In non-dimensional form, normalizing time by the surface-tension time-scale $t_\sigma = \sqrt{\rho h_0^3/\sigma}$, the inviscid equations read  
\begin{align}
    &\partial_t v+v v_z=-p_z \qquad & p = \frac{1}{h\sqrt{1+h_z^2}}-\frac{h_{zz}} {(1+h_z^2)^{3/2}} \label{eq:eom1}\\
    & \partial_t h+v h_z=-v_z\frac{h}{2} \label{eq:eom2}\\
     & h\left(t, \frac{\pm L(t)}2\right)=1 \qquad &  v\left(t,\frac{\pm L(t)}2\right)=\pm \sqrt{\text{We}}
\label{eq:eom3}
\end{align}
where $v(z)$ is the axial velocity of the air inside the bridge, the non-dimensional bridge length is $L(t)=L_0+2t\sqrt{\text{We}}$, and the non-dimensional speed of the boundary is $v_p=\sqrt{\text{We}}$. Partial derivatives with respect to $z$ are denoted by a subscript. The momentum balance appears in equation (\ref{eq:eom1}), with inertia on the one hand and the pressure difference due to surface tension, proportional to the mean curvature, on the other. Viscous effects are neglected (Appendix~\ref{sec:modeling}). Equation ~(\ref{eq:eom2}) represents conservation of mass $\propto h^2$ inside the bridge. 
We use pinned boundary conditions for the height in our simulations, Eq.~\ref{eq:eom3}, since we do not observe significant slippage in experiments (Appendix~\ref{sec:slip}). We solve Eqs.~\ref{eq:eom1}-\ref{eq:eom3} numerically (Appendix~\ref{sec:numerics}).

\section{Main results: satellite sizes}
Our first main result is that the volume of the satellite bubble, formed as a by-product of the breakup, increases with We and $L_0$, as can be seen in Fig.~\ref{fig:images}$(a)$ from the experimental data (the same trends are found in simulations).
We quantify the increase in size of the satellite in Fig.~\ref{fig:images}$(c)$ where the satellite volume normalized by the initial volume, $V_b/V$, is plotted as a function of We for three values of $L_0$. For each $L_0$ and We number we performed 10 experiments, the results of which fall on top of each other as seen in Fig.~\ref{fig:images}$(c)$. Thus, the volume of the final bubble is extremely reproducible and robust. In particular, special care to control the airflow around the bridge or have the same precise velocity profile for the plates is not required to reproduce the same values. 

\begin{figure*}[ht]
    \begin{centering}        \includegraphics{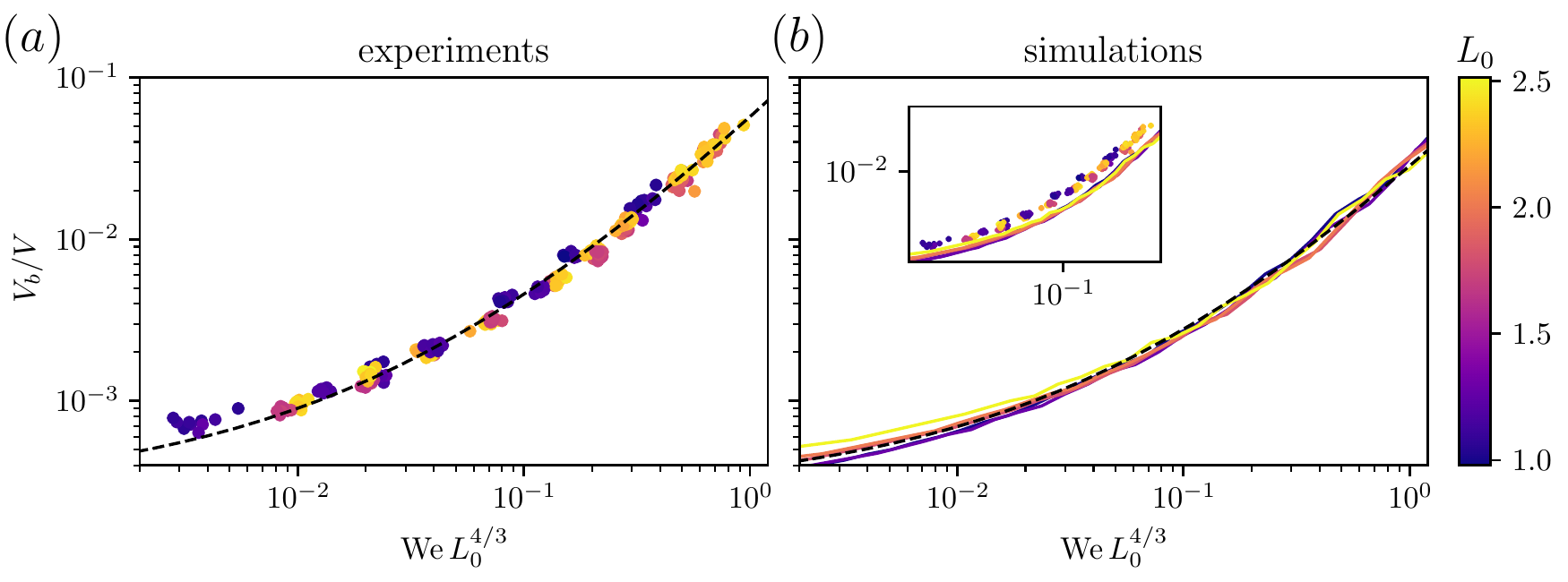}
        \caption{The fractional bubble volume, $V_b/V$ as a function of ${\rm We}\,L_0^{4/3}$ in $(a)$ experiments, and $(b)$ simulations. The colors denote different initial aspect ratios $L_0$, and the dashed lines represent our theoretical formula, Eq.~\ref{eq:V_formula}. The inset shows the general agreement between experiments (symbols) and simulations (lines).
        \label{fig:Vb}}
    \end{centering}
\end{figure*}

Denoting by $h_c$ the radius at the mid point of the bridge when it is flat, as shown in Fig.~\ref{fig:images}$(b)$, we find that $V_b\approx 13 h_c^3$ in simulations and experiments, demonstrated in Fig.~\ref{fig:images}$(d)$ where lines show results from simulations and dots show an ensemble average over experimental results. Thus, $h_c$ is a single characteristic length scale determining the final satellite volume. Furthermore, for $h_c$ we find the simple linear dependence: 
\begin{equation}
   h_c=h_c^0+a \sqrt{\text{We}}\, L_0 
   \label{eq:h_c}
\end{equation}
as shown in Fig.~\ref{fig:images}$(e)$. The values of $h_c^0$ and $ a$ can be obtained from a linear fit of $h_c$ as a function of $\sqrt{\text{We}}L_0$. We find that $a\approx 0.2$ ($a\approx 0.15$ in simulations), while $h_c^0=A L_u(L_0)$ with $A\approx 0.014$ in simulations and experiments. 
As described below, $L_u(L_0)$ is the critical bridge length  for a bridge of volume $L_0$  beyond which there is no equilibrium bridge shape \cite{gillette_stability_1971}.

Therefore, $V_b/V$ is captured by the detailed functional form:
\begin{equation}
\begin{split}
    \frac{V_b}V &= \frac{13\left(A L_u+a\sqrt{\text{We}}\, L_0\right)^3}{\pi L_0}
\\
&=\frac{13}\pi \left(A L_u L_0^{-1/3}+a \sqrt{\text{We}\, L_0^{4/3}}\right)^3
    \end{split}
    \label{eq:V_formula}
\end{equation}
The factor $L_u(L_0)L_0^{-1/3}$ can be directly computed from the known equilibrium shapes. While it generally depends on $L_0$, for the range considered here the variation is not large, and, since $A/a$ is small, will be apparent only for $\text{We}\, L_0^{4/3}<10^{-2}$.
Indeed, the fractional final volume $V_b/V$ 
roughly collapses as a function of the combined parameter $\text{We}\, L_0^{4/3}$, as shown in Fig.~\ref{fig:Vb}$(a)$ for the experiments and in Fig.~\ref{fig:Vb}$(b)$ for the simulations. A slight spread with $L_0$ is observable for simulations with $\text{We}\, L_0^{4/3}<10^{-2}$. 
In the same figures, formula \eqref{eq:V_formula} is shown as a dashed line for $L_0=1.8$, using the corresponding inferred constants $a$ and $A$. It captures the satellite volume across roughly two orders of magnitude extremely well, both in experiments and simulations.  

Below we will offer an explanation for these findings, suggesting that formula \eqref{eq:V_formula} results from three dynamical stages. Initially, the bridge evolves quasi-statically: its shapes closely matching a sequence of static equilibrium shapes (Sec.~\ref{sec:static}). When the aspect ratio reaches $L_u$, an instability at the center of the bridge develops (Sec.~\ref{sec:Stage_I}). We will show that the duration of this stage is independent of $\rm We$ and is proportional to $L_0$. The length of the central region is thus increased by a factor proportional to $\sqrt{\text{We}}\, L_0$ during this stage, due to the bridges stretching. In the third stage, the dynamics follow a universal evolution controlled by a single length scale, proportional to $h_c$ (Sec.~\ref{sec:stage_2}). These two observations combine to give Eq.~\ref{eq:V_formula} for the final satellite volume, as described in Section~\ref{sec:scaling}. We summarize and discuss possible implications of our findings in Sec.~\ref{sec:summary}.

\begin{figure*}[ht]
    \begin{centering}        \includegraphics{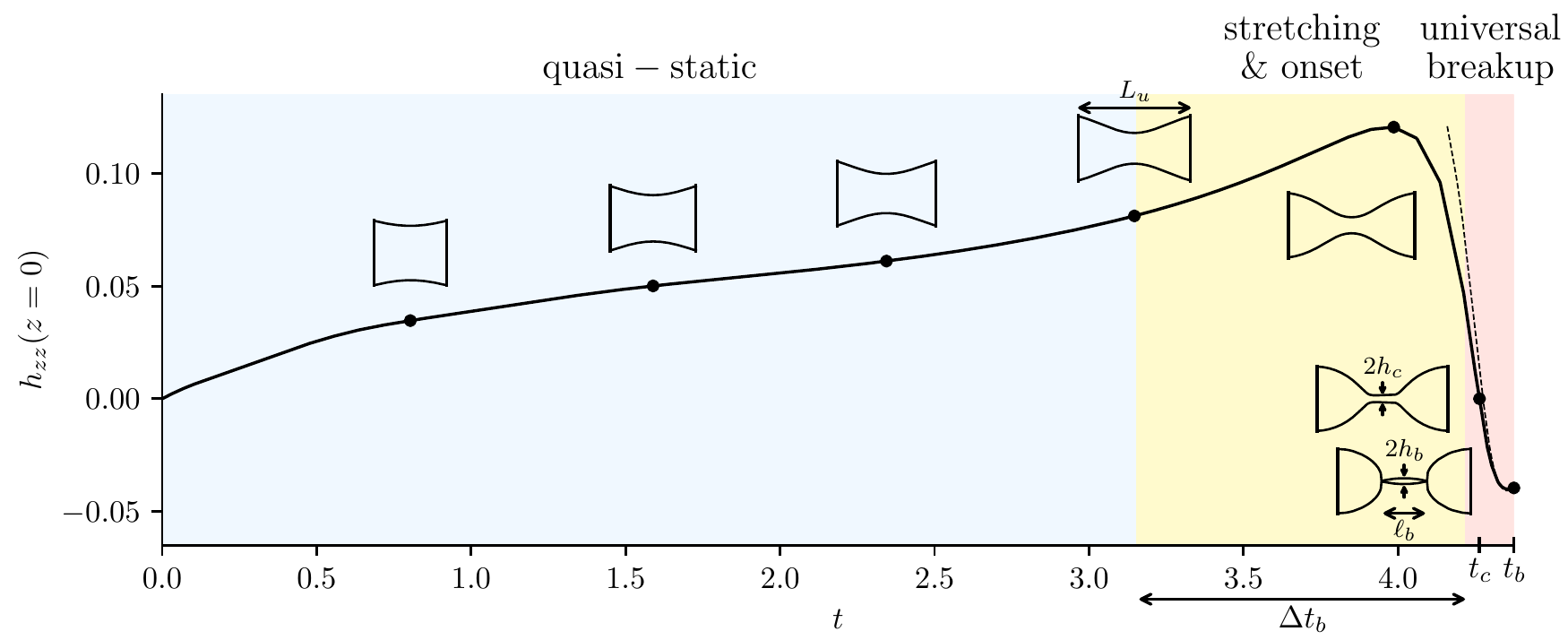}
        \caption{Sketch of stages of the dynamics. Data is taken from the evolution of a simulated liquid bridge with ${\rm We}=0.07$ and $L_0=1.8$; all variables are non-dimensionalized with $h_0$ and the surface-tension time-scale. The line shows $h_{zz}$ at the center of the bridge; at representative points denoted with dots, we plot the full shape of the bridge. In the first stage of evolution (blue), the bridge follows a sequence of equilibrium shapes. The unstable equilibrium is reached once $L(t)=L_u$. This begins the second stage of evolution (yellow) which lasts for $\Delta t_b=\mathcal{O}(1)$. In the second stage, the central region of the bridge stretches as the breakup process begins. Finally, the central region approaches a universal breakup trajectory, which is the third stage of evolution (red). The universal trajectory is plotted as a dashed line.
        \label{fig:evolution}}
    \end{centering}
\end{figure*}

\section{Stages of the breakup dynamics}
\label{sec:dynamic_stages}
As the bridge stretches, it undergoes three distinct phases prior to fragmentation: quasi-static dynamics, stretching and instability onset, and universal breakup dynamics, as summarized in Fig.~\ref{fig:evolution}. 
\subsection{Quasi-static dynamics}
\label{sec:static}
Our experiments and simulations are in the regime $\text{We}\ll 1$, where the plate separation is slow compared with the surface-tension time scale.
Thus, as the length of the bridge $L(t)$ slowly increases with time, the bridge takes a surface-area minimizing shape.
Such constant mean curvature shapes have been extensively studied in the past~\cite{gillette_stability_1971,lowry_capillary_1995}. In particular, for pinned boundaries, zero Bond number and a conserved volume $V$, there is a unique stable, connected equilibrium shape for each pair $V,L$ up to a critical aspect ratio $L_u(V)$ (Appendix~\ref{sec:theory}).
We show the close agreement between the shapes in simulations and the sequence of equilibrium bridge shapes when $L\lesssim L_u$ in Appendix~\ref{Appendix:dynamics}.
At $L(t)=L_u$ the equilibrium solution becomes unstable, while for $L>L_u$ a constant mean curvature solution with the given volume no longer exists~\cite{gillette_stability_1971}.
Quasi-static dynmaics ends once the bridge reaches this critical length $L_u(L_0)$.
This is typically the longest dynamical phase.

\begin{figure*}[t]
    \begin{centering}
    \includegraphics{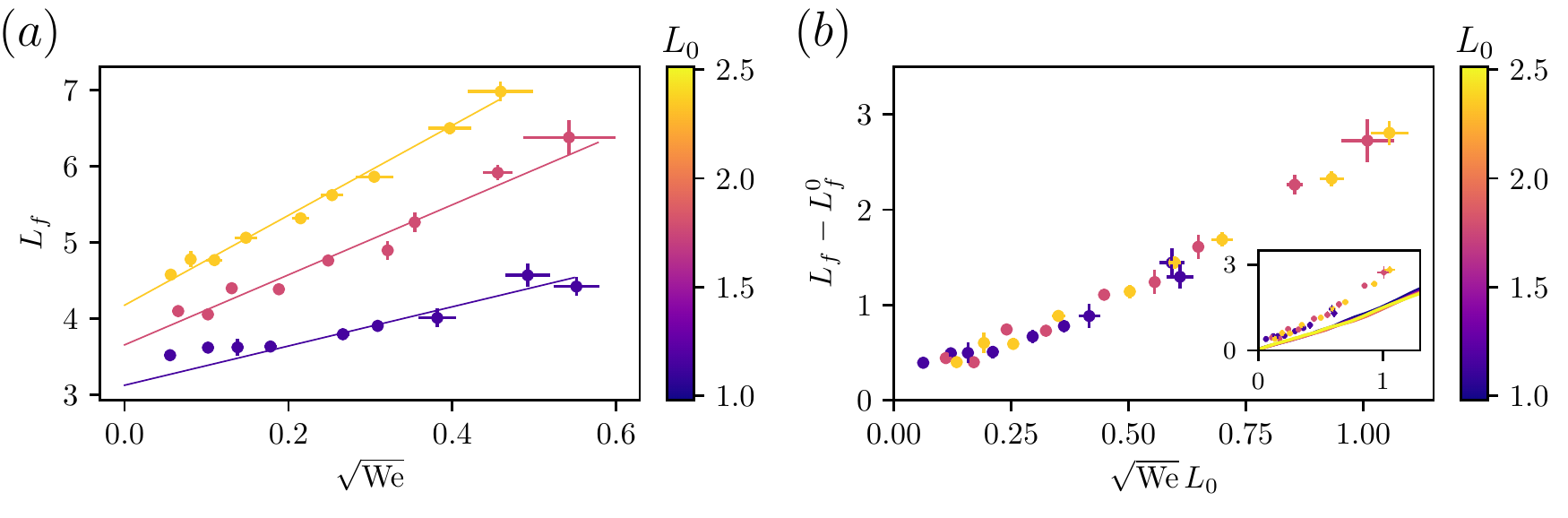}
    \caption{$(a)$ The final bridge length $L_f$ as a function of $\sqrt{{\rm We}}$. We find $L_f$ is linearly proportional to $\sqrt{{\rm We}}$, but with a slope and constant offset $L^0_f$, extracted through a linear fit, both of which depend on $L_0$. Dots show the ensemble average of experimental results with error bars representing one standard deviation. $(b)$ $L_f-L_f^0$ as a function of $\sqrt{\rm We} \, L_0$ for experiments with different initial aspect ratios $L_0$. The inset includes simulation data, depicted with lines for different $L_0$. While $L_f-L_f^0$ is linearly proportional to $\sqrt{\rm We}\, L_0$ in both experiments and simulations, the constant of proportionality is larger in experiments than in simulations.
    \label{fig:stretch}}
    \end{centering}
\end{figure*}

As breakup starts at $L(t)=L_u$, the length of the bridge at pinch-off, $L_f$, approaches the critical value $L_f\to L_u(L_0)$ in the limit ${\rm We}\to 0$.
In Fig.~\ref{fig:stretch}$(a)$ we plot
$L_f$ as a function of $\sqrt{\text{We}}$ for the experimental data (the same dependence is observed in simulations), where dots represent an ensemble average over experiments and error bars show one standard deviation. A linear fit is overlaid, and experiments with different $L_0$ cross the y-axis at different points, corresponding to $L_f^0\equiv L_f(\text{We}=0)$.
As expected, we find $L_f^0\approx L_u$ (see Appendix \ref{Appendix:dynamics}).

\subsection{Stretching and instability onset}
\label{sec:Stage_I}
Once $L(t)= L_u,$ the bridge becomes unstable and starts to breakup.
In this stage, the bridge stretches before it is attracted to the universal breakup solution (described in the next section).
This stretching phase starts when $L(t)=L_u$, and ends when the central region of the bridge becomes flat (e.g., Fig.~\ref{fig:images}$(b)$ and Fig.~\ref{fig:evolution}). 

We find the duration of this stretching phase, $\Delta t_b$, is independent of We, and increases with $L_0$.
As the bridge is stretched at a constant velocity $2\sqrt{\text{We}}$, and the duration of the final breakup phase is very short, we expect that $L_f\approx L_f^0+2\sqrt{\text{We}}\Delta t_b$.
We find that $L_f-L_f^0$ increases linearly with $\sqrt{\rm We}$, as shown in Fig.~\ref{fig:stretch}$(a)$, implying the time it takes for the bridge to break $\approx \Delta t_b$, is independent of ${\rm We}$.
We also notice that experiments with different $L_0$ have a different slope, i.e. different $\Delta t_b$. In particular, Fig.~\ref{fig:stretch}$(b)$ shows that $\Delta t_b\propto L_0$, giving the functional dependence $L_f=L_f^0+b\sqrt{\text{We}} \, L_0$. 

The same features are observed in simulations, presented alongside experiments in the inset of Fig.~\ref{fig:stretch}$(b)$.  
Note that the slope $b$ differs between simulations and experiments (Fig.~\ref{fig:stretch}$(b)$). We obtain $b=2.4$ in experiments and $b=1.3$ in simulations, implying this stage is twice as long in the experiments than in the simulations. This may be related to differences in the bridge dynamics near the boundaries. However, these details do not alter the scaling of the satellite volume with $\sqrt{\text{We}}\, L_0$, which is the robust feature of interest here. 

We have demonstrated that the bridges breakup time is unaffected by the stretching of the bridge. More generally, the dynamics are similar to that of a soap film bridge between two open rings held at its critical length \cite{chen_dynamics_1997}. Initially, the fluid accelerates away from the necking point due to the pressure imbalance, and both the advection term and the inertia induced by the stretching boundaries are negligible. As in Ref.~\cite{chen_dynamics_1997}, this process of fluid acceleration continues until advection from the central region becomes as important as pressure gradients. The inertia induced by stretching, which is a small effect when $\text{We}\ll 1$, is then subdominant compared to the inertia that has developed due to the instability, as we demonstrate in Appendix~\ref{sec:theory}.
As advection becomes important, fluid starts to escape the entire central region, causing the height to flatten, corresponding to the start of the universal breakup stage. In this next stage the surface will overturn, sealing a satellite. 

While the onset of the instability appears to be insensitive to the stretching of the bridge, stretching does enter in two important ways. First, close to the boundaries of the bridge the induced inertia cannot be neglected. The induced We-dependent flow is thus negligible only in a finite region around the bridge center, corresponding to the breakup region, whose extent we denote by $z_b$.  Second, the stretching of the bridge  changes the bridges length, thus changing the local curvature, and so the pressure, everywhere---including in the central region, see Appendix \ref{sec:theory}. Our finding that the duration of the breakup is roughly independent of We is therefore surprising, and is a feature that remains to be explained.

\begin{figure*}[t]
    \begin{centering}\includegraphics{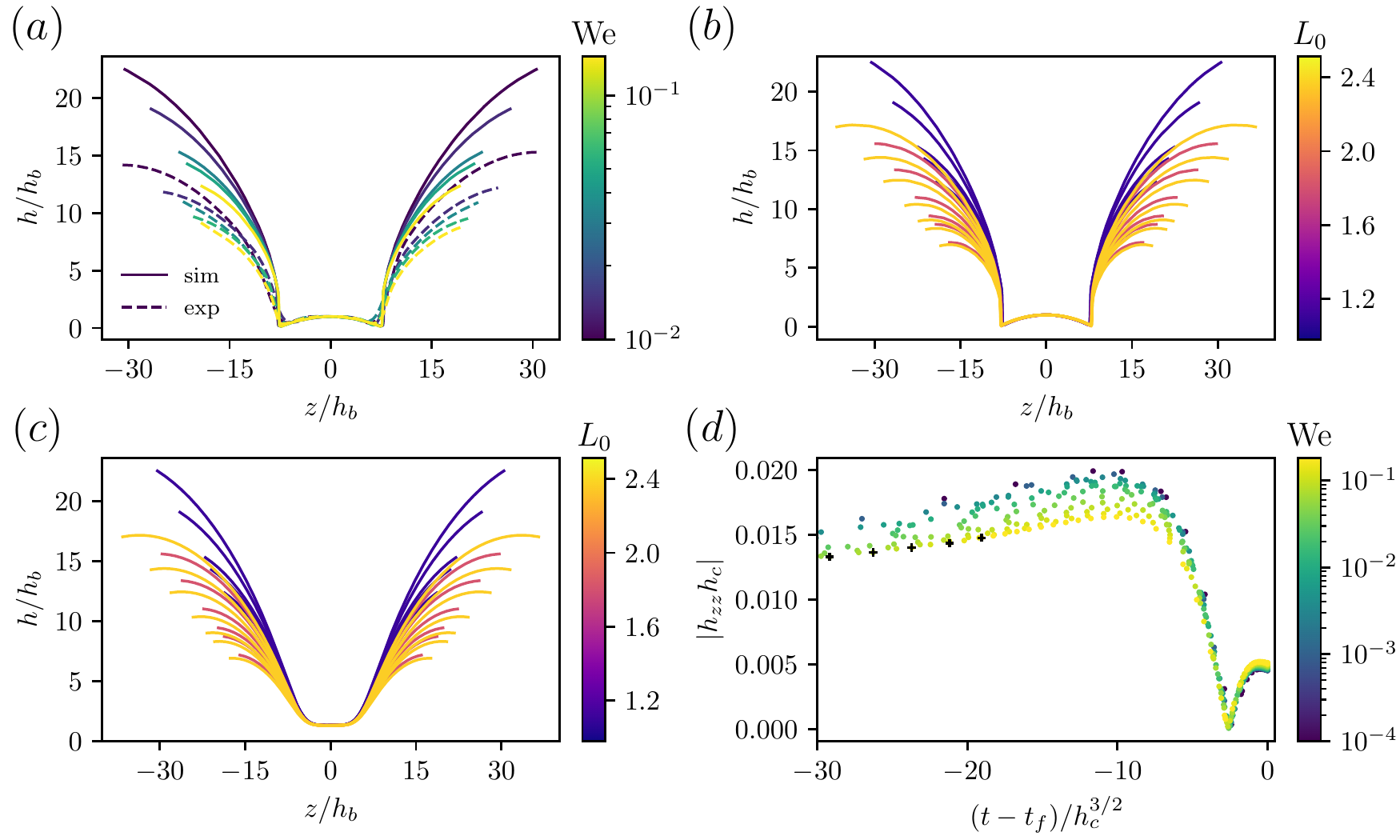}
    \caption{$(a)$ Bridge shape at the pinch-off time in experiments (dashed lines), and simulations (solid lines), for $L_0\approx 1.1$ and various ${\rm We}$. When normalized by $h_b$, the bridge shapes of the experiments and simulations at different ${\rm We}$ collapse near the central region. $(b)$ Bridge shapes in simulations at the pinch-off time, with different $L_0$ (colors) and ${\rm We}$ (high curves correspond to lower ${\rm We}$). The central region of the bridges collapses to a universal shape independent of $L_0$ and ${\rm We}$. $(c)$ Bridge shapes in simulations at $t_c$, the time at which $h_{zz}(z=0)=0$, with different $L_0$ and ${\rm We}$. $(d)$ $h_{zz}(z=0)$ as a function of time, both normalized using $h_c$. Dots represent the trajectories of different simulations with $L_0\approx 1.1$, colored by their ${\rm We}$. For the highest ${\rm We}$, we plot a black cross at the time of the last stable equilibrium when $L(t)=L_u$. While simulations with different ${\rm We}$ initially have different curvatures, they are attracted to a universal breakup solution prior to $t_c$ where $h_{zz}=0$.}
    \label{fig:self-sim}
    \end{centering}
\end{figure*}

\subsection{Universal breakup dynamics} 
\label{sec:stage_2}
The dynamics in the central region of the bridge (away from the plates) become universal towards the end of the breakup.
We first demonstrate the universality of the shape of the bridge in the central region at the pinch-off time $t_f$. In Fig.~\ref{fig:self-sim}$(a,b)$, we plot the height $h(z)$ at $t_f$, with both lengths rescaled by $h_b$, the height at the center of the bridge at this time. We see a very good collapse of the shape in the central region, independent of both $L_0$ and We.
The collapse is only seen in the central region, not near the boundaries.
While there is good agreement between experiments and simulations in the central region in Fig.~\ref{fig:self-sim}$(a)$, the bridges take quite different shapes outside this region. 

To show that the collapse is not simply dictated by properties of the shape at the final time, in Fig.~\ref{fig:self-sim}$(c)$ we present the shape of the bridge at the time $t_c$ when the surface flattens ($h_{zz}(t_c,0)=0$, see Fig.~\ref{fig:evolution}).  
This is the beginning of this universal phase. We again rescale both axes by $h_b$; the collapse of the central region is evident. In particular, as expected $h_b\propto h_c=h(t_c,z=0)$.

Using results from numerical simulations, we now wish to demonstrate that the breakup dynamics follow a universal time-dependent trajectory.
That is, a finite region forms near the center of the bridge where, if both time and length-scales are properly rescaled, the height and velocity dynamics follow a universal trajectory independent of We and $L_0$. We have already demonstrated that the central region of the bridge takes a universal form when rescaled by the length-scale $h_c$, at times $t_c$ and $t_f$. Taking $h_c$ as the relevant length scale and assuming that breakup is driven by surface tension, from dimensional analysis we expect the temporal scaling to be $(t_f-t) \propto h_c^{3/2}$. This is the well known surface tension-inertia scaling, as in Ref.~\cite{eggers_singularities_2015}, based on the length scale $h_c$. In Fig.~\ref{fig:self-sim}$(d)$ we plot the curvature at the center of the bridge in such rescaled coordinates. Indeed, under this rescaling, trajectories with different We all asymptotically converge to the same universal curve. We observe that simulations with higher We approach this curve later in (rescaled) time, but that by the time $t_c$ ($h_{zz}=0$), all our simulations follow the universal dynamics. 
While the scaling $t-t_f \propto h_c^{3/2}$ is not new, to the best of our knowledge this is the first time it is observed for a universal solution occupying a finite, constant in time, spatial region. Previous studies of inertia-surface tension driven breakup focused on the pinch-off points and the similarity dynamics there~\cite{lister_capillary_1998,chen_dynamics_1997,eggers_singularities_2000, eggers_singularities_2015,Casterjon_plethora_2015}.

We now discuss why universal dynamics emerge in this system. The universal breakup phase occurs quickly, i.e., much faster than the surface tension time $t_\sigma$ (see Fig.~\ref{fig:evolution}). Thus, the length of the bridge is roughly constant during this phase (equivalent to ${\rm We}=0$). Next, we estimate the typical height of the central breakup region to be $h_c$, which is much smaller than the height of the bridge at the boundaries $h_0=1$. 

We consider the ansatz $h=h_c H\left[\frac{t_f-t}{t_c},\frac{z}{z_c}\right]$ and $v=v_c V\left[\frac{t_f-t}{t_c},\frac{z}{z_c}\right]$ in equations~(\ref{eq:eom1}) \& (\ref{eq:eom2}), and require that all terms are of the same order in $h_c\ll 1$. We expect that in the breakup region the inertia and surface tension terms are comparable, although We based on the plate velocity is assumed to be small. That is because the pressure imbalance, due to the instability, drives a fast flow with a local ${\rm We}\sim\mathcal{O}(1)$. Using the ansatz that all terms in (\ref{eq:eom1}) \& (\ref{eq:eom2}) are of the same order, we find that the horizontal scale $z_c\propto h_c$. In addition, $v_c\propto z_c/t_c$ and, as expected from an inertia-surface tension balance, $t_c\propto h_c^{3/2}$, in agreement with our findings in Fig.~\ref{fig:self-sim} (a more careful analysis, showing that the bridges length $L\approx L(t_f)$ does not enter, is presented in Appendix~\ref{sec:theory}). 
Note that this temporal scaling is consistent with our assumption that the bridge is hardly stretched during this stage: since $(t_f-t)\propto h_c^{3/2}\ll \Delta t_b = O(1)$, the change in $L$ is negligible.

We thus obtain the following universal form
\begin{equation}
    \begin{split}
    &h(t,z)=h_c H\left[\frac{t_f-t}{h_c^{3/2}},\frac{z}{h_c}\right] \equiv h_c H[T,Z]\\
   & v(t,z)=h_c^{-1/2} V\left[\frac{t_f-t}{h_c^{3/2}},\frac{z}{h_c}\right]\equiv h_c^{-1/2}V[T,Z]
   \label{eq:scaling}
    \end{split}
\end{equation}
with the functions $H[T,Z]$ and $V[T,Z]$ satisfying equations~(\ref{eq:eom1})-(\ref{eq:eom2}), which do not depend on any external parameters. External parameters can enter through the boundary conditions, but
the outlined considerations apply only to the breakup region, and so the relevant boundary conditions are to be taken at $z=z_b$. To determine those we assume that $z_b/h_c\gg 1$ so that in terms of the variable $Z$ the boundary is at infinity. We require that the solution in the breakup region match the external solution, where we assume that the velocity is finite, i.e. that $v(t,z_b)$ does not scale with $h_c$. We thus get the boundary condition $ V\left[T,Z\right]= v(t,z_b)/v_c=h_c^{1/2}v(t,z)\to 0$ as $Z\to\infty$. Next, assuming that the height of the external solution has a finite curvature, i.e. that $\partial_{zz} h(t,z_b)$ is independent of $h_c$, we get the condition $\partial_{ZZ} H[T,Z]\to h_c\partial_{zz} h(t,z_b)=0$ as $Z\to\infty$. The matching is thus to be done at the inflection point of the external solution. Equivalently, this implies the boundary condition $H[T,Z]/Z\to O(1)$ as $Z\to\infty$. To summarize, under these assumptions, the boundary conditions for $H,V$ which replace (\ref{eq:eom3}) are
\begin{align}
    H[T,Z\to \infty]\propto  Z  && V\left[T,Z\to\infty\right]= 0
    \label{eq:bc_H}
\end{align} 
Equations~(\ref{eq:eom1})-(\ref{eq:eom2}) which $H,Z$ satisfy together with the boundary conditions~(\ref{eq:bc_H}) are seen to be universal, independent of external parameters. The breakup of the bridge should be captured by the attracting dynamical solution of this system, explaining our findings in Fig.~\ref{fig:self-sim}.

\section{The satellite bubble volume}
\label{sec:scaling}
Here we combine our insights from the previous section to explain the dependence of the volume of the satellite bubble on We and $L_0$ as shown in Fig.~\ref{fig:Vb} and quantified in Eq.~\eqref{eq:V_formula}.
We expect for the volume of the satellite that $V_b\propto h_b^2 l_b$, where $h_b$ is the height at the center of the bridge, and $l_b$ the lateral distance between the pinch-off points at the final time (see Fig.~\ref{fig:evolution}). In Figs.~\ref{fig:self-sim}$(a)-(c)$ we have shown that $l_b\propto h_b\propto h_c$, explained by the universal breakup solution (Eq.~\ref{eq:scaling}). Thus, in agreement with our observations in Fig.~\ref{fig:images}$(d)$, $V_b\propto h_c^3$, with a universal constant of proportionality. 

We next explain the linear dependence of $h_c$ on $\sqrt{\rm We}$ and $L_0$ (Eq.~\eqref{eq:h_c}).
As $l_b$, the length of the final bubble is proportional to $h_c$, we will equivalently argue why $l_b$ has a linear dependence on $\sqrt{\rm We}$ and $L_0$.
Without stretching, starting with a bridge of critical length $L_u$, we expect for the satellite length to be proportional to the total bridge length $l_b^0\propto L_u$. This motivates the relation
\begin{equation}
    h_c^0=A\, L_u
    \label{eq:A}
\end{equation}
with $A$ independent of initial conditions and $L_0$.

For finite We, the breakup region extends during the stretching phase (dynamical stage II), so we have $l_b = l_b^0 + 2 v_b \Delta t_b$, where $v_b\approx v_p=\sqrt{\rm We}$ is the stretching velocity of the breakup region (Appendix~\ref{Appendix:dynamics}).
Since $h_c\propto l_b$, and $\Delta t_b\propto L_0$, we recover the linear dependence of $h_c$ on $\sqrt{\rm We}$ and $L_0$ in Eq.~\ref{eq:h_c}.
Our experiments and simulations have slightly different values of $a$, the proportionality constant between $h_c$ and $\sqrt{\rm We}\, L_0$, which is due to differences in the breakup time and the stretching velocities of the breakup region.
Put together, these correspond to values of $a$ which are 25\% smaller in the simulations than the experiments. Combining equations \eqref{eq:A} and \eqref{eq:h_c} together with the finding $V_b=13h_c^3$ we recover Eq.~\eqref{eq:V_formula}.
The differences in $a$ between experiments and simulations lead to satellite volumes which are smaller by $\sim 50\%$ in the simulations (inset of Fig~\ref{fig:Vb}(b)).
Crucially, these differences do not alter the properties of the universal dynamics or the scaling of the satellites volume.

\section{Summary and discussion}
\label{sec:summary}
A fluid bridge that is held between two plates and slowly stretched  will eventually break, undergoing a dramatic topological change. 
We find that, when the bridge is stretched at constant speed, the outcome of the breakup is highly predictable. A small satellite bubble always forms, its volume uniquely determined by two non-dimensional parameters: the (normalized) volume $L_0$ of the bridge and the normalized stretching speed $\sqrt{\text{We}}$. By varying these two parameters, the volume of the satellite bubble can be made to vary over two orders of magnitude, reaching up to $10\%$ of the total bridge volume. 

The high reproducibility of the bubble size can be explained by the universality of the breakup dynamics, revealed by our simulations and experiments. In the slow stretching regime $\text{We}\ll1$ considered here, the stretching has a negligible effect on the breakup dynamics itself, which are spatially localized to the central part of the bridge. Instead, the dramatic dependence of the satellite bubble size on We is a consequence of the stretching of the breakup region, occurring prior to the universal breakup solution. Correspondingly, we find a simple dependence of the bubble size on the combination $\sqrt{\text{We}}\, L_0$ which arises due to two robust features. First, the universal breakup solution depends on a single length scale determining the bubble size. Second, it takes a fixed time, independent of the stretching and proportional to $L_0$, to approach the universal solution (starting from a bridge with the unstable equilibrium length $L_u(L_0)$). It is mainly during this time interval that the breakup region is stretched. The bubble size can thus be increased either by increasing $L_0$, which prolongs the time over which the breakup region is stretched prior to the convergence to the universal solution, or by increasing We, increasing the stretching rate of that region during the approach. 

Our results also shed light on the universal features of static breakup, i.e., the breakup dynamics of a fluid bridge of a fixed critical length $L_u(L_0)$. In particular, while the universality of pinch-off dynamics is a well established phenomena, e.g.~\cite{eggers_singularities_2015}, the novelty here is that the attracting universal solution spans a finite region in space, remaining finite even as pinch-off is approached. Thus, it is not a similarity solution. Instead, we find that the extent of the breakup region is proportional to $L_u(L_0)$, with a universal proportionality constant. What selects this constant, determining the disconnected solution that the universal dynamics converges to (and thus the satellite size), remains to be understood. The dependence of the duration of the breakup process on the liquid bridge volume $L_0$, is another robust aspect of the breakup process which 
merits future investigation.

Our work demonstrates that the singular nature of breakup can lead to universality of satellite formation, going beyond the universality of pinch-off dynamics. This opens the possibility of universality of satellite formation driven by other force balances---as is the case for pinch-off---an exciting prospect for understanding satellite formation and drop-size distributions in turbulent flows~\cite{riviere_capillary_2022}.     

\section*{Acknowledgements}

The authors thank Albane Smilga for inspiring this project, and Bruno Le Floch for the introduction. We thank Howard Stone for generously hosting us in his lab, and a critical reading of this manuscript. We also thank numerous members of the Stone lab for their generous help: Janine Nunes, Ching-Yao Lai, Suin Shim and Maksim Mezhericher. We gratefully acknowledge Jens Eggers for useful conversations and Kinneret Keren for helpful comments on this manuscript. A.F. would like to thank Yuval Elhanati and Eldad Afik for advice on the image analysis.  A.F.~and D.L.~are supported by PCTS fellowships, and BSF grant No. 2022107. A.F.~is supported by ISF grant No. 486/23. D.L.~is supported in part by NASA HTMS
grant 80NSSC20K1280 and NASA OSTFL grant 80NSSC22K1738. Computations were conducted with support by the NASA High End Computing Program through the NASA Advanced Supercomputing (NAS) Division at Ames Research Center on Pleiades with allocation GID s2276.

\appendix

\section{Experiments}
\label{sec:experiment}

\subsection{Experimental setup}
\label{sec:exp-setup}
For our experiments, we form bubbles using a soap solution, using a 10:1 mixture of distilled water and blue Dawn dish soap. The thickness of the soap film in our experiment was smaller than $500\, {\rm nm}$ (blue colors could be observed in the interference patterns on our soap films).
Using a syringe we form a bubble of a fixed volume and place it between two wetted acrylic plates.
To achieve the desired initial aspect ratio, we slowly pull the plates apart to form a cylindrical soap bubble.
During this process, the contact points slip, corresponding to a fixed contact angle with the plates of $\pi/2$.
For these Neumann boundary conditions, the maximum initial aspect ratio is $L_0=\pi$.

To drive the bubble to breakup, we pull apart the acrylic plates using a motor.
The plates are glued onto two pillars which are mounted onto a low friction track.
The pillars are pulled by strings which are connected to an electric motor through a pulley, converting the rotation of the motor to linear motion of the pillars and plates. The speed of the plates is controlled by changing the input voltage to the motor.
We use a high speed camera to capture the breakup process, using a frame-rate in the range of $4000-6500$ frames per second, depending on We.
The breakup process takes $\sim 30-300\, {\rm ms}$.

\subsection{Data analysis}
\label{sec:exp-image}

We analyze the images with the scikit-image Python package~\cite{van_scikit_2014}, using the canny edge detector to extract the soap-film height and the final bubble volume. The height is computed as the half-distance between the upper and lower free surface of the bridge. The final bubble volume is computed after the bridge breaks, once the bubble relaxes to a spherical form. Then, for a series of images, we compute the bubble area and extract its radius. The volume is then computed based on the average (over the series of frames) radius. We compute the velocity $v_p(t)$ from the soap-film bridge length $\frac{L(t)}2$. We fit $\frac{L(t)}2$ using a cubic spline (through the UnivariateSpline scipy package~\cite{SciPy-NMeth}) and then take its derivative.

We estimate $t_c$, the time at which the curvature at the center of the bubble is zero, by finding the first time the curvature in the central region is negative. We define the central region using a We-dependent height threshold.
There are sometimes oscillations of the location of the minimum before the curve overturns at the center, which typically leads to an overestimation of $h_c$. 
To mitigate the effects of these oscillations on our analysis, we also require that in the next image the curvature in the same region remains negative. Still, we believe that these oscillations bias the determination of $h_c$, leading to the lower values of $V_b/h_c^3$ seen in Fig.~\ref{fig:images}$(d)$ for We in the range $0.006-0.03$ (where we empirically observe such oscillations).

\section{Numerical methods}
\label{sec:numerics}
To solve Eqns.~\ref{eq:eom1}-\ref{eq:eom3}, we change coordinates to $(s,\tau)$, where $s=z/(\frac{L(t)}2)$ and $\tau=\log(L(t)/L_0)=\log(1+2t/L_0)$ (note this is a different definition of $\tau$ than used in Appendix~\ref{sec:theory}), and $t$ is non-dimensionalized with the advection timescale. Then the equations of motion become
\begin{align}
    &\partial_\tau v - s \partial_s v + {\rm We}^{-1} \partial_s p = -v\partial_s v \nonumber \\
    &\partial_\tau \log h - s \partial_s \log h + \frac{1}{2}\partial_s v = -v\partial_s \log h \nonumber \\
    &p=\frac{e^{\tau}}{h\sqrt{e^{2\tau}+\left(\frac{2}{L_0}\right)^2h_s^2}} - \left(\frac{2}{L_0}\right)^2\frac{e^{\tau}h_{ss}}{\left[e^{2\tau}+\left(\frac{2}{L_0}\right)^2h_s^2\right]^{3/2}} \nonumber \\
     & h(\tau,\pm 1)=1 \qquad  v(\tau, \pm 1)=\pm 1
\label{eq:numerics}
\end{align}  
We solve the equations of motion with the Dedalus pseudospectral framework \cite{burns2020}.
We represent $v$, $\log h$, $h_s$, and $p$ as Chebyshev series with 1024 modes.
Nonlinear terms are calculated using the 3/2 dealiasing rule, although the presence of more complicated nonlinear products introduce some aliasing errors.
We initialize with $p=\log h=1$, $h_s=0$, and $v=\tanh\left[(2.2/\pi)\tan(\pi s/2)\right]$.
We run simulations for $2/L_0$ ranging from $0.8$ to $2.0$ in increments of $0.5$, and for ${\rm We}=10^{-4},$ $10^{-3}$, and 30 values ranging from $10^{-2.5}$ to $10^{0.5}$, equally spaced in logarithm.

To timestep the equations, we use implicit-explicit timestepping methods from \cite{Ascher97}.
In particular, we use the 2nd order, two-stage Runge--Kutta method for the ${\rm We}=10^{-4}$ simulations, and the 3rd order, four-stage Runge--Kutta method for all other simulations.
All terms linear in the perturbation variables are treated implicitly, and all nonlinear terms are treated explicitly.
As the bridge approaches breakup, the equations become more stiff.
To improve the stability of the timestepping scheme, we define a base state, and evolve the (fully nonlinear) deviation of each variable away from this base state.
This allows us to implicitly timestep more linear terms in the equations than evolving the variables directly.
We set the base state to be given by the problem variables every 100 timesteps.
We initially take timesteps of size $10^{-4}$, and evolve the system until breakup.
To get better time resolution near breakup, we then restart from the simulation state between 150-200 timesteps before the breakup, taking timesteps that are half as large.
We continue to iteratively restart the simulations with smaller and smaller timesteps until the timestep size is $6.25\times 10^{-8}$.
For ${\rm We}=10^{-4}$, we start with timesteps of size $2\times 10^{-4}$, and iteratively restart the simulations until the timestep is $1.25\times 10^{-7}$.

\section{The three dynamical stages}
\label{Appendix:dynamics}
\subsection*{Quasi-static bridge evolution}

\begin{figure*}
\centering
\includegraphics{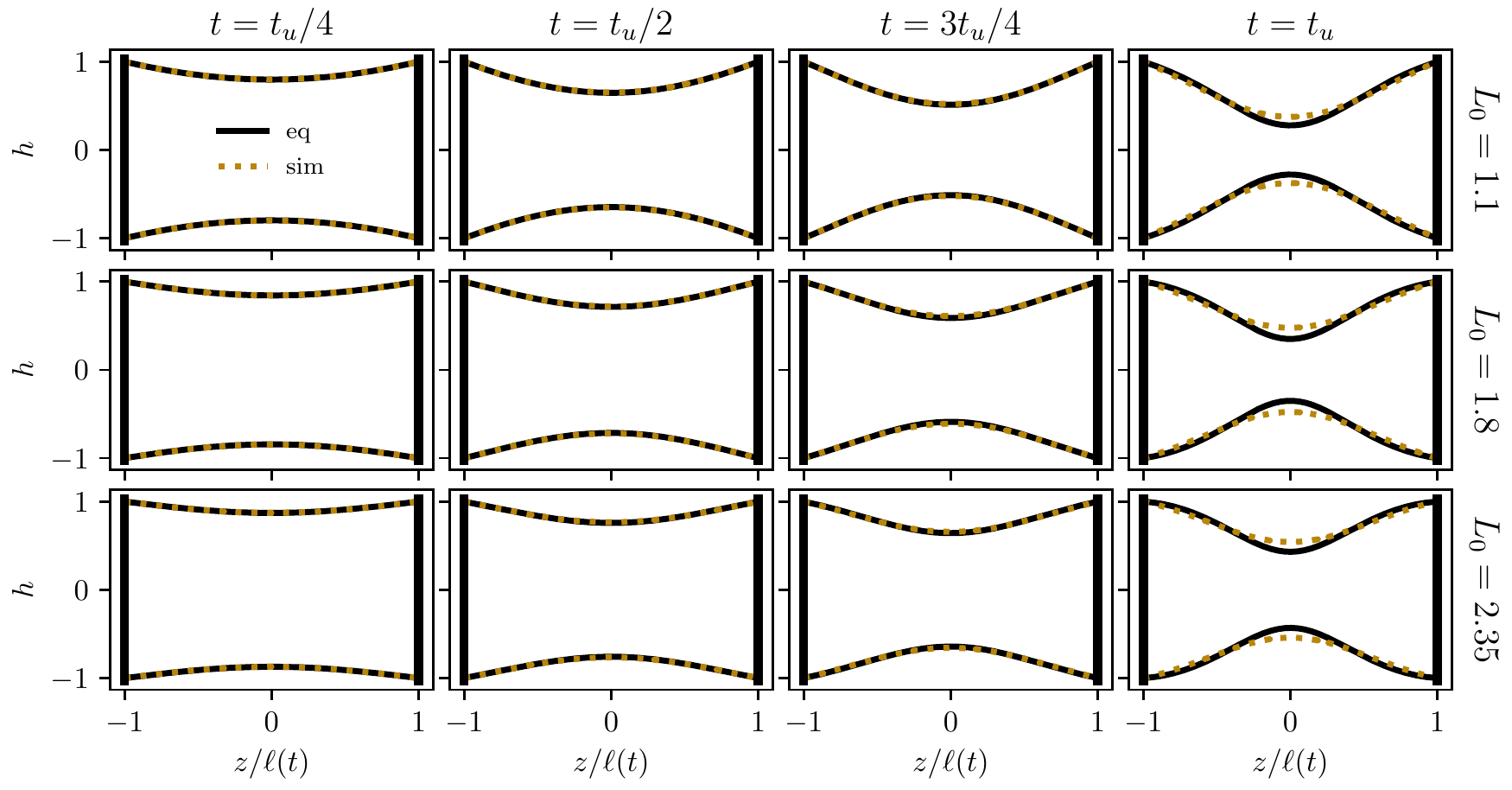}
\caption{Equilibrium bridge shapes (black lines) and simulated bridge shapes (dotted yellow lines) for ${\rm We}=0.07$. We include bridges with different initial aspect ratios, and at different times.}
\label{fig:SI_Leq}
\end{figure*}

In Fig.~\ref{fig:SI_Leq} we plot equilibrium bridge shapes along with simulated bridge shapes in simulations with ${\rm We}=0.07$. The simulated bridges are very close to the equilibrium bridges, until we approach $t_u$, the time at which the equilibrium bridge becomes unstable. As ${\rm We}$ decreases, the quasi-static dynamics follow the equilibrium shapes more closely.

\begin{figure}[ht!]
\centering
\includegraphics[width=0.45\textwidth]{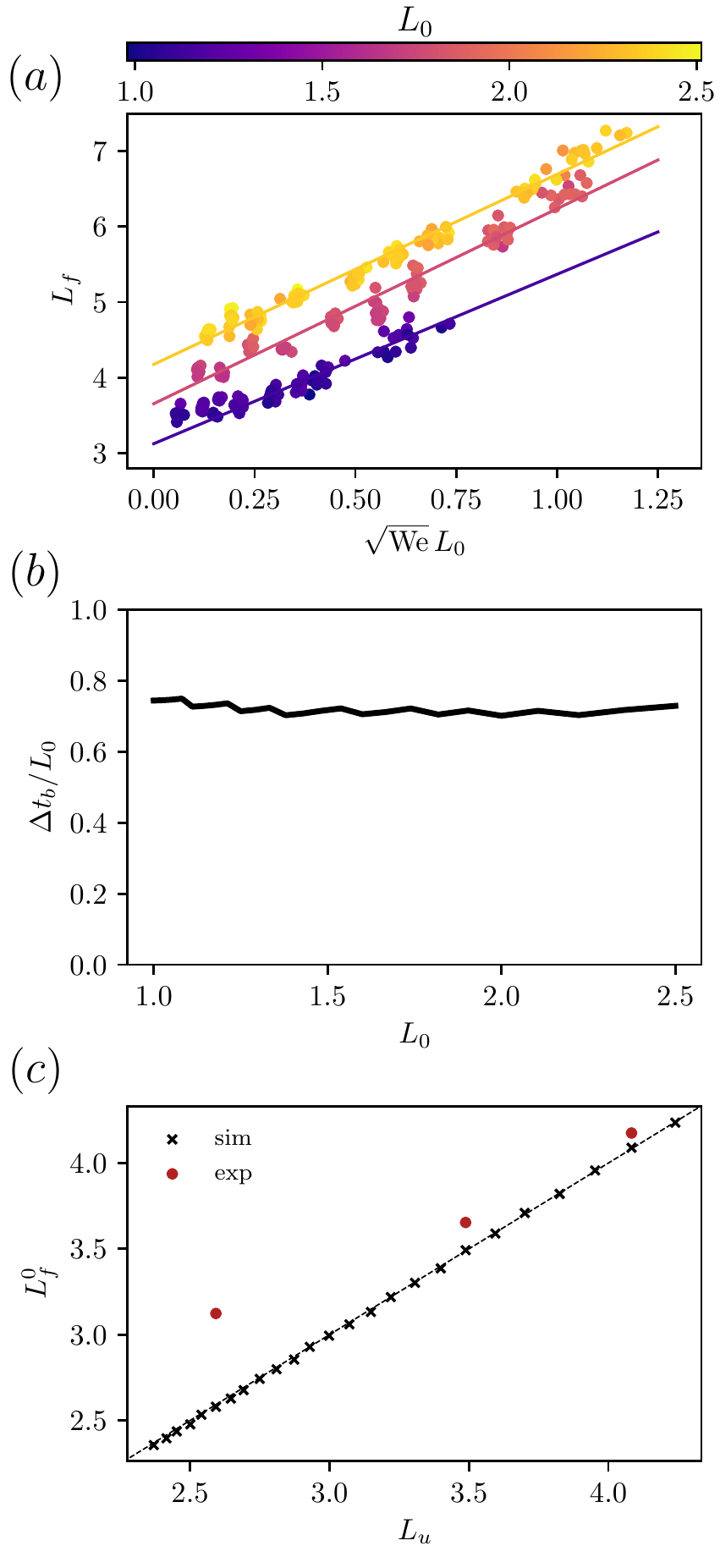}
\caption{$(a)$ The soap-film bridge length at breakup $L_f$ (experiments) and the linear fit from which we extract $L^0_f$ and $\Delta t_b$ in the main text.
$(b)$ $\Delta t_b/L_0$ as a function of $L_0$ as extracted from a linear fit$L_f=L_f^0+2\sqrt{\text{We}}\Delta t_b$ in the simulations.
$(c)$ $L_0^f$ as a function of $L_u(L_0)$. Crosses show simulation results, red dots show experimental results. The dashes line shows the relationship $L_0^f=L_u$, which is very well satisfied in the simulations.}
\label{fig:L_f}
\end{figure}

\begin{figure*}
\centering
\includegraphics[width=\textwidth]{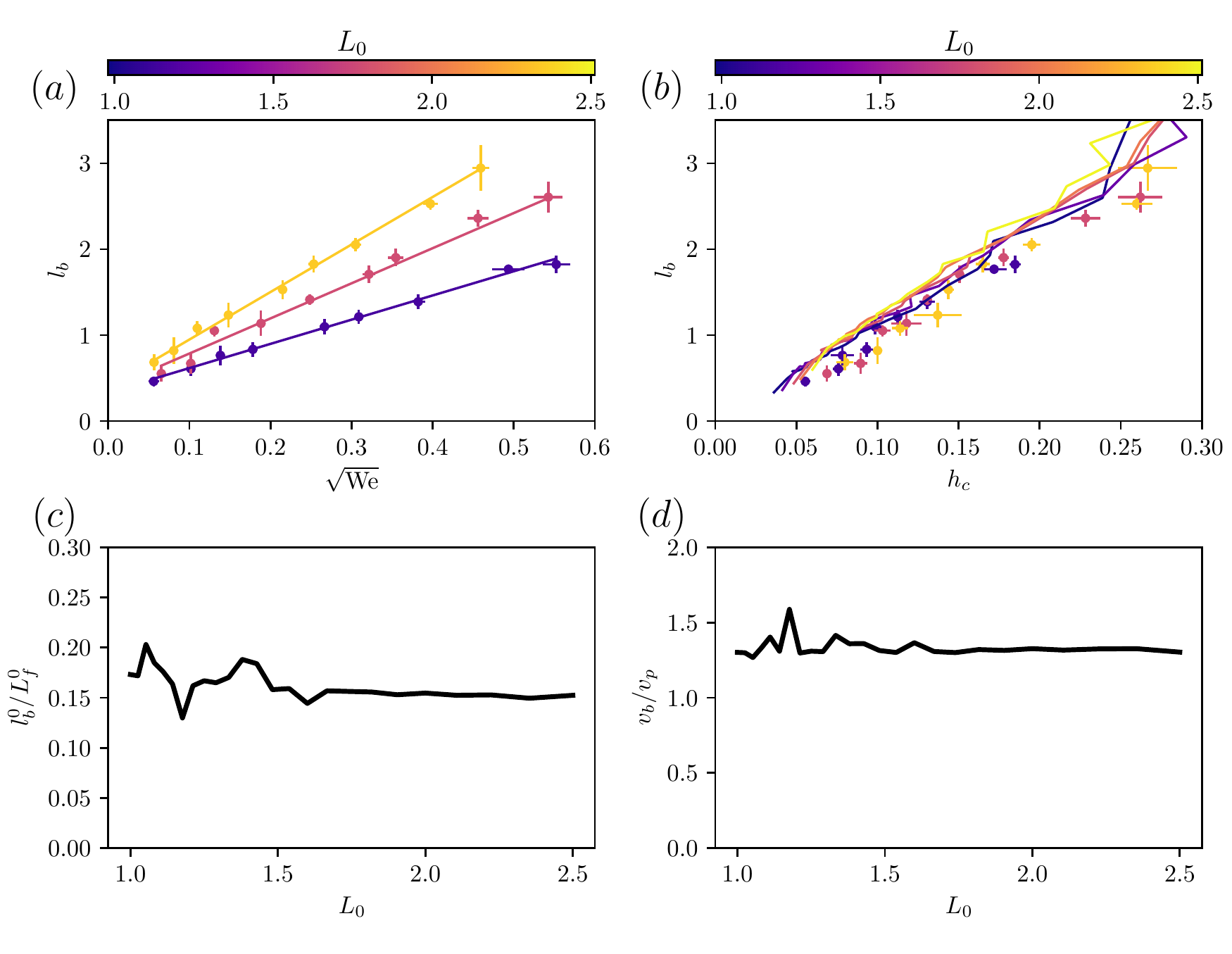}
\caption{$(a)$ The satellite length at breakup $l_b$, as a function of $\sqrt{\text{We}}$ and the linear fit used in the main text (experiments). $(b)$ $l_b$ as a function of $h_c$ for experiments and simulations. $(c)$ $l^0_b/L^0_f$ as a function of $L_0$ as extracted from linear fits in the simulations. ($d$) The ratio of speeds of the breakup region stretching and that of the plates, $v_b/v_p$, as a function of $L_0$, extracted from the slope in the linear fit of $l_b$ and $\Delta t_b$.}
\label{fig:l_b}
\end{figure*}

\subsection*{Final bridge length $L_f$ and time to breakup}

We show the bridge length at breakup $L_f$ as a function of $\sqrt{\text{We}} \,L_0$ in Fig.~\ref{fig:L_f} $(a)$, where each point represents a single experiment and the lines show the linear fit. In Fig.~\ref{fig:L_f} $(b)$ we show how the breakup timescale $\Delta t_b/L_0$, extracted from the linear fit $L_f=L^0_f+2\sqrt{\text{We}}\Delta t_b$ for each simulation, varies with $L_0$. Fig.~\ref{fig:L_f} $(c)$ shows $L_0^f$, extracted from this fit, as a function of $L_u$, the unstable bridge length. In the simulations, $L_0^f$ is very close to $L_u$. In the experiments, $L_0^f$ is somewhat larger, likely due to slippage of the bridge contact line for low ${\rm We}$ experiments (Appendix~\ref{sec:slip}).

\subsection*{The characteristic scale in the universal regime: $l_b$ and $h_c$}
In Fig~\ref{fig:l_b} we present the scaling of $l_b$, the lateral distance between pinch-off points at the breakup time of the bridge. In panel $(a)$ we show the ensemble averaged data from experiments as a function of We, along with a linear fit to $l_b=l_b^0+c\sqrt{\text{We}}\, L_0$. It can be seen that the different lines indeed have different slopes corresponding to the different $L_0$. In panel $(b)$ we show that the data collapses when plotted as a function of $h_c$ and that there is a linear dependence between $l_b$ and $h_c$. In this panel, the dots correspond to experimental results, and the lines correspond to simulation results.

Panel $(c)$ shows the ratio $l_b^0/L_f^0$ as a function of $L_0$ for the simulations, showing it is roughly constant. In panel $(d)$, $v_b/v_p\equiv(l_b-l_b^0)/(L_f-L_f^0)=c/b$ is shown as a function of $L_0$ for our range of simulations using the linear fits. The ratio is very near to being constant, equal to $1.3$.

\section{Theoretical considerations}
\label{sec:theory}

\subsection{Equilibrium shapes}
\label{sec:eq}
The equilibrium bridge shapes are completely characterized by two parameters, which are typically chosen to be $L = 2l/h_0$, where $2l$ is the length of the bridge, and $S=V/V_{cyl}$ where $V$ is the volume of the bridge, and $V_{cyl}$ is the volume of a cylindrical bridge with length $L$ and height $h_0$.
The parameters $S$ and $L$ are not independent during the dynamics: $S(t)= l(0)/l(t) = L_0/L(t)$. Note, that we always start from an initial cylindrical bridge, i.e $S(0)=1$. We use the parameterization introduced in~\cite{gillette_stability_1971} to determine the equilibrium shape for each $(L,S=L_0/L)$ pair (solving the corresponding equations numerically using Mathematica). For each initial aspect ratio $L_0$, we determine the critical bridge length $L=L_u$ once these equations no longer have a solution.

\subsection{Inertia driven double pinch-off in a symmetric set-up}
Let us first rewrite equations~(\ref{eq:eom1})-(\ref{eq:eom3}) in a coordinate system $(s,\tau)$ where the inertia from stretching of the bridge is more evident: $z= \frac{L(t)}2 s, \quad \tau=t$ so that $\partial_s = \frac{L(t)}2\partial_z, \quad \partial_\tau=\partial_t+s\sqrt{\text{We}}\partial_z$.
Using the modified velocity $u=v-s\sqrt{\text{We}}$, we arrive at
\begin{align}
    &l(\tau)\partial_\tau u+u(\partial_s u+\sqrt{\text{We}})=-p_s \nonumber\\ & p = \frac{l(\tau)}{h\sqrt{l(\tau)^2+h_s^2}}-\frac{l(\tau)h_{ss}} {(l(\tau)^2+h_s^2)^{3/2}} \nonumber \\
    & l(\tau)\partial_\tau h+u \partial_s h=-\frac{h}{2}(\partial_su+\sqrt{\text{We}} )\nonumber\\
     & h(\tau,\pm 1)=1 \qquad &  u(\tau, \pm 1)=0
\label{eq:stretched}
\end{align}  
with $l(\tau)\equiv\frac{L(\tau)}2=\frac{L_0}2+\tau\sqrt{\text{We}}$.

We see that if $p_s\sim u\partial_s u$ while $\partial_s u\gg \sqrt{\text{We}}$ (equivalently $\partial_sv =l(t)\partial_z v\gg \sqrt{\text{We}}$), the inertial term would no longer be negligible but the explicit dependence on We would still disappear from the equations.
Indeed, this is what happens at the end of the second stage, as a fast outflow develops in the central region (note the flattening of the free surface of the bridge) implying $\partial_z v\gg1$ there.
On the other hand, this condition is not satisfied close to the boundaries: from equation~(\ref{eq:eom2}) and the boundary conditions~(\ref{eq:eom3}), the condition $\partial_s u =-\sqrt{\text{We}}$ should be dynamically satisfied there (reflecting that the surface is pinned at the boundaries so that the radial velocity should remain zero there). Note that once the fluid inertia becomes important, in the limit $\partial_s u\gg \sqrt{\text{We}}$, also $u\approx v$ so that at leading order we can use $u=v$ in the breakup region. Thus we do not make a distinction between $u$ and $v$ in our discussion of the universal solution in the main text.

Here we derive the universal breakup dynamics more carefully, taking into account the length of the bridge $l(\tau)\approx l(\tau_f)$ as an additional parameter (which depends on We and $\rm L_0$) entering the dynamics, as seen in equations~(\ref{eq:stretched}). 
Similarly to the main text, we use the ansatz $h=h_c H\left[\frac{t_f-t}{t_c},\frac{s}{s_c}\right]$ and $u=u_c U\left[\frac{t_f-t}{t_c},\frac{s}{s_c}\right]$. Note that here the anzatz is in the $s$ coordinate, which is normalized by the length of the bridge, as apposed to $z$. Plugging it into equation~(\ref{eq:stretched}) with $\text{We}=0$, and requiring that all terms are of the same order in the limit $h_c\ll1$, we get that $h_c=l s_c=z_c,$ $v_c=u_c=ls_c/t_c=z_c/t_c$ and $t_c=h_c^{3/2}$. As compared to the main text, $l$ appears explicitly in the dynamical equations. Then, the equations for $H, U$ become universal in the limit $h_c\ll1$, i.e. independent of $l$, only if one assumes $l s_c=h_c$. This gives the relation we use in the main text $h_c=z_c$, resulting in the scaling form~(\ref{eq:scaling}) for $H,U\equiv V$, which then satisfy equations~(\ref{eq:eom1})-(\ref{eq:eom2}), where $l$ no longer appears.

\section{Modeling}\label{sec:modeling}

\subsection{Compressibility, viscosity and gravitational effects} The compressibility of the air inside the bridge can be neglected. The largest Mach number based on the plate velocity is $\text{Ma} = v_p/c_s \sim 10^{-2}$.
The flow velocity induced by surface tension will be at most an order of magnitude larger since $\text{We}\sim 0.1$ in those experiments and $h_c\sim0.1 h_0$.
Thus, the largest Mach number based on the flow velocity will still be $\ll 1$.
Furthermore, the pressure difference induced by surface tension is at most of the order of $\sigma/h_{min} \sim 10^2\, {\rm N}/{\rm m}^2 $ (taking $h_{min} = 0.05 h_0\approx 10^{-4}\,{\rm m}$), while the atmospheric pressure is $1\,\text{atm} = 10^5\, {\rm N}/{\rm m}^2$, so that the deviations in the density of the air are of the order $10^{-3}$.
Equivalently, we may compare the scaling of the pressure gradients in the leading order $\propto \text{Ma}^{-2}$ to the stress induced by surface tension $\propto \text{We}^{-1}$ in the momentum balance equation, giving $ \text{Ma}^{-2}\gg  \text{We}^{-1}$, i.e., the flow is incompressible in the leading order. 

We measure the surface tension of our soap solution to be $\sigma = 25\, {\rm mN}/{\rm m}$. It is measured using the shape of a pendant-drop, using a matlab code developed for this purpose in the Howard Stone lab. Viscous stresses become comparable to surface tension stresses at the scale $l_\nu = \rho \nu^2/\sigma=5 \times 10^{-9}\, {\rm m}$ where $\nu$ is the kinematic viscosity of air $\nu \approx 1.5\times 10^{-5} \, {\rm m}^2/{\rm s}$. The maximal Ohnesorge number based on the minimal height $h_c\sim 1 \, {\rm mm}$ in the experiment is given by $\text{Oh} =\sqrt{l_\nu/h_c} $ and is of the order of $\text{Oh} \sim 10^{-3}$. The Ohnesorge number at the pinch-off points does eventually become $O(1)$, so that viscous effects become important. This however does not influence the earlier breakup stages in the central region of the bridge on which we focus here.   
In addition, the minimum Reynolds number in the experiments is $\text{Re} = \sqrt{\text{We}} /\text{Oh} \sim 10^2$ implying that viscous effects are sub-leading both to inertia and surface tension during the breakup and before pinch-off. 

The maximal Bond number is $\text{Bo} = g\rho h^2/\sigma \approx 0.05$ so that the effects of gravity are negligible (as evidenced by the axisymmetry of the soap-film bridge up to breakup).

\subsection{The soap-film surrounding the bridge}
The inertia of the liquid in the soap film, compared to that of the air inside the bridge, may be neglected in our experiments for most of the breakup process. To compare the inertia of the two we assume that the acceleration of the two fluids is comparable, meaning that we only need to compare their masses. The thickness of the soap film is smaller than $500\, {\rm nm}$, so we estimate $h_s\sim 1\, \mu\text{m}$. The density of the soap solution is approximately that of water, $\rho_s \sim 10^3\, {\rm kg}/{\rm m}^3$. The condition on the height of the bridge such that the soap solution inertia is negligible is thus: $2\pi h_{min} h_s \rho_s \ll \pi h_{min}^2 \rho_a$, giving $h_{min} \gg 2\rho_s/\rho_a h_s \sim 10^{-3} \, \text{m}\approx 0.1 h_0$. 
The smallest bubble heights occur in the central region before breakup.
These are smallest for the lowest We, reaching heights of $0.05 h_0$.
Thus, our lowest We experiments may be somewhat affected by soap film inertia in the final stages of breakup.

\subsection{The one dimensional approximation}
\label{sec:1D}
We model the dynamics by the equations of a slender jet, even when the aspect ratio $L$ is not large. Initially, the justification for this approximation is that deviations from a radially constant axial velocity $v_z(r,z,t) = v(z,t)$ and pressure $p(r,z,t)=p(z,t)$ are small: the initial pressure is uniform (constant atmospheric pressure and constant mean curvature for equilibrium shapes), and the boundary conditions for the velocity (which drives the dynamics) are also uniform in the radial direction. Then, working in cylindrical coordinates with $0\leq r\leq h(z,t)$, we can expand around a radially constant solution for $v_z\approx v_z^0(z,t)+\tilde{v}_z(r,z,t)$ and the pressure $p(r,z,t)=p_0(z,t)+\tilde{p}(r,z,t)$, assuming deviations $\tilde{v}_z, \tilde{p}$ to be small in amplitude. The radial velocity is then determined by incompressibility and is given in the leading order by $v_r = -v_z^0 r/2 $, and in particular at the free surface $v_r = -v_z^0 h/2$. This is the same relation obtained within the slender jet approximation, which means that $v_z^0$,  $p_0(z,t)$ and $h(z,t)$ satisfy the equations used in the slender jet approximation~\cite{eggers_drop_1994} (i.e the momentum equations for $v_r, v_z$ are self consistent to leading order, so that radial perturbations do not grow). In addition, when the bridge is stretched radial perturbations become more and more suppressed, as it enters into the regime of a slender jet.

\subsection{Differences between experiments and modeling}
There are several physical aspects in which the detailed dynamics in experiments and simulations are distinct. One difference is that the stretching protocol is not the same: while in simulations the edges are pulled with a constant speed throughout, in experiments that speed invariably changes in time, potentially with asymmetries between the two plates. Another difference is in the boundary conditions. There is some slippage of the edges of the bridge in experiments while those are pinned in simulations (Appendix~\ref{sec:slip}).

\section{Experimental Considerations}
\subsection{Contact line slippage}\label{sec:slip}
\begin{figure*}
\centering
\includegraphics[width=\textwidth]{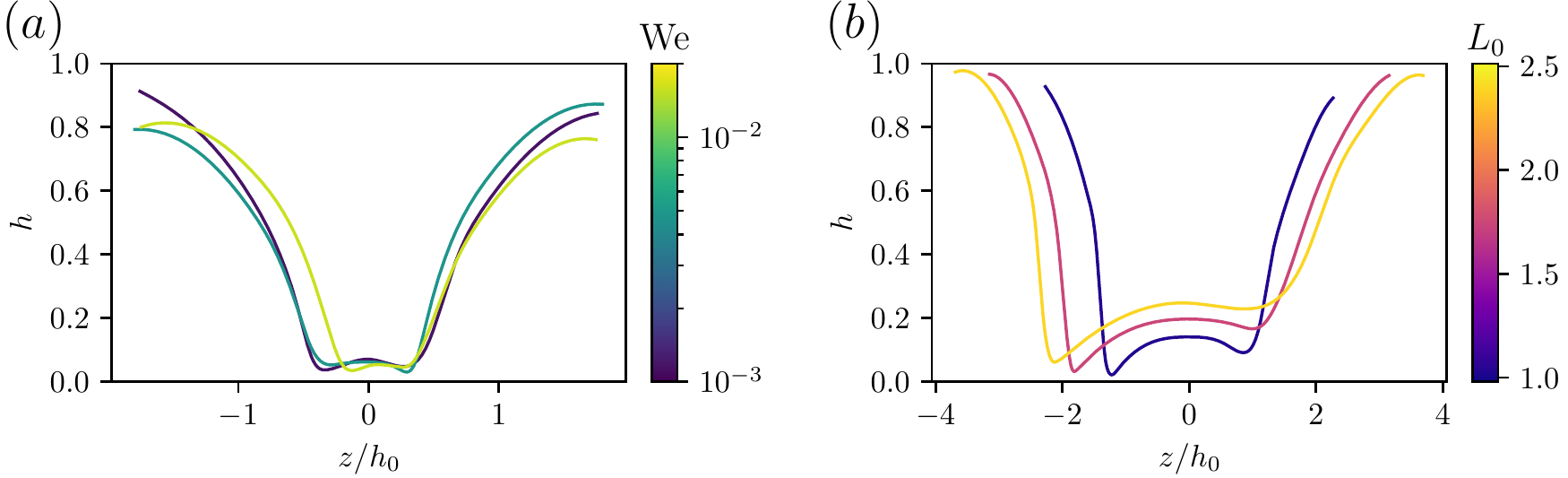}
\caption{$(a)$ The soap film bridge shape at breakup for $L_0=1.1$. Note that the bubble height at the boundaries is lower than $1$, implying there was slippage at the boundaries. $(b)$ Asymmetry in the breakup shape for the highest We for three different aspect ratios, $(L_0, {\rm We})\approx (1, 0.48)$, $(1.7, 0.35)$, and $(2.4, 0.24)$. The asymmetry is inherited from the asymmetry between the two plates.}
\label{fig:low_hi}
\end{figure*}

Most of our experiments were performed rapidly enough so that the soap film bridge effectively remained pinned at the boundaries. However, for the smallest aspect ratio we used, $L_0\sim 1.1$, and especially for the slowest experiments, $\text{We}\sim 0.002$, we did observe slippage of the soap film at the boundaries. Indeed,  we observe an $\sim 20\%$ slippage of the contact line of the soap film bridge at the plates for the lowest We. We observe this phenomenon for all the bridges with these parameters, an example is given in Fig.~\ref{fig:low_hi} $(a)$ where the shape of bridges with three different We is shown.
This indicates that the timescales forstretching and slipping are similar in these experiments, probably due to the long time it takes to reach $L_u$ for this $L_0$ ($L_u$ being more than twice larger than $L_0$). The main effect of slippage is to increase the actual normalized volume $L_0$ for these experiments (since we use the height at the plates to normalize the volume), pushing $L_u$ to higher values. We therefore discard the three lowest We experiments for the inference of $L^0_f$. Despite this, we find  the difference  between the expected $L_u$ and $L_f^0$ to be the largest for this aspect ratio. 

\subsection{Definition of We}
\begin{figure*}
\centering
\includegraphics{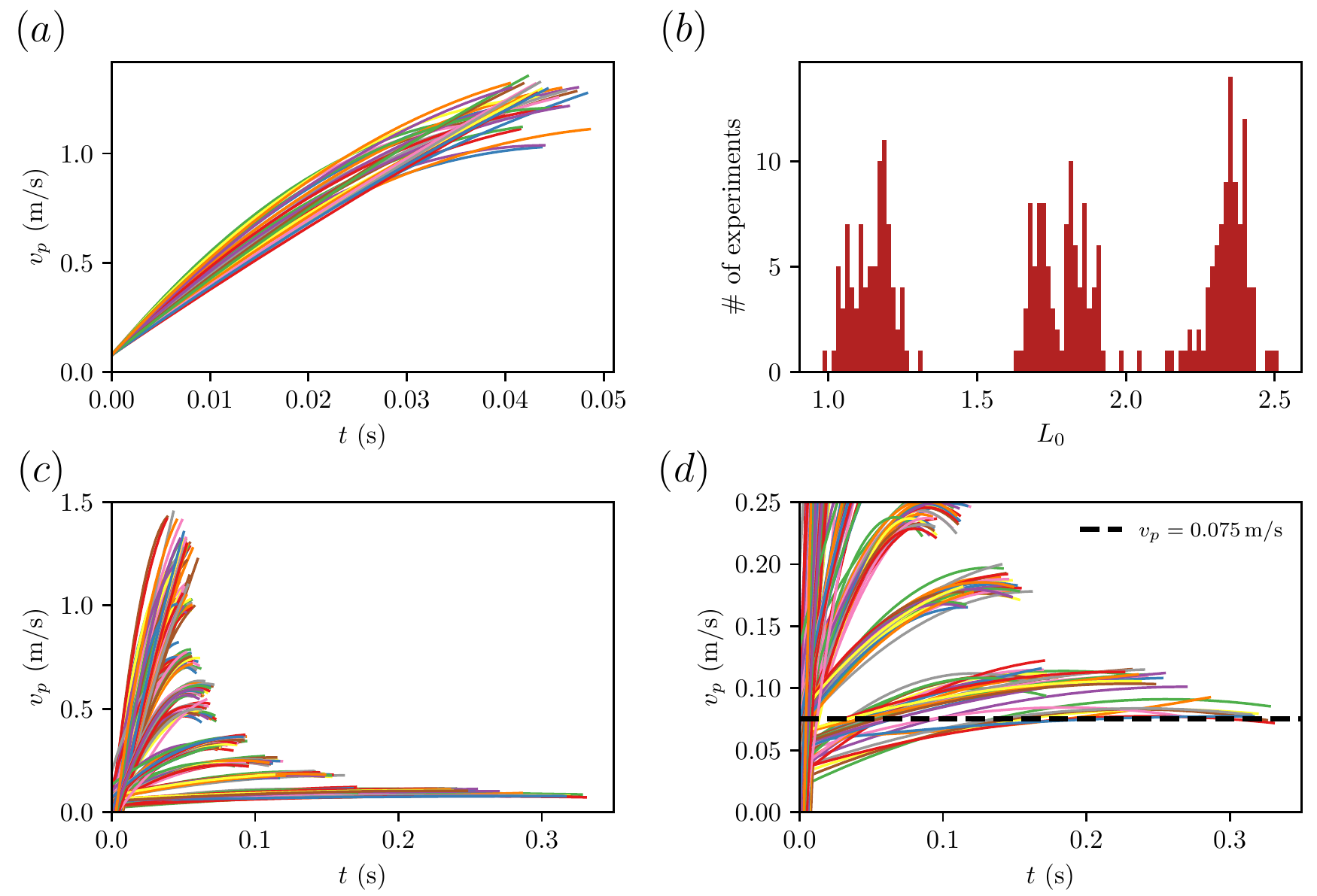}
\caption{$(a)$ Velocity profile for a particular motor rotation speed, producing We-number in the range $1.6-2.5$. $(b)$ Histogram of the initial aspect ratio of the soap-film bridge in the experiments. $(c)$ Velocity profiles for the different experiments. $(d)$ Zoom in on small velocities and the threshold used for the definition of We. }
\label{fig:velocity}
\end{figure*}

To define the Weber number, we must determine the velocity of the plates for each experiment. In the experiments the velocity changes with time, as the speed with which the bridge is stretched increases from zero until it roughly saturates. We present the corresponding velocity profiles (as extracted from $l(t)$) in Fig.~\ref{fig:velocity}. To define We for each experiment we compute the time-averaged square velocity, beginning from a threshold speed equal to $0.15\, {\rm m}/{\rm s}$, so that the acceleration phase is not included, and more weight is placed on the later stages (when the relevant dynamics are expected to occur). In Fig.~\ref{fig:velocity} $(d)$ we show how that threshold compares with the slowest experiments. Introducing such a threshold is physically justified for the purposes of the analysis in the main text since we expect the early stages to correspond to the quasi-static phase of the dynamics, which is irrelevant for the breakup dynamics in the later stages. However, it is also true that while We plays a key role in the analysis in the main text, it is not uniquely defined in the experiments (since the plate velocity does not quite plateau). We have tried different definitions (such as using the averaged velocity or the velocity when the bridge length increases by a particular, large enough, increment) which did not produce a significant difference for our results except for a slight multiplicative increase in We. Note that the discrepancy between experiments and simulations could not be resolved by such an increase, as the required shift in We would be too large, corresponding to plate velocities which are larger than the maximal plate velocity in the experiment (also, a single such correction could not produce a match between experiments and simulations for both $L_f$ and $V_b/V$). Finally, note the ratio of the plate velocity and the stretching velocity of the breakup region $v_b/v_p$ as inferred for the experiments is independent of such multiplicative factors in the definition of We.

\subsection{Symmetric breakup in experiments and role of acceleration}
Two important features of our pulling protocol are that the pulling of the two plates is roughly symmetric (but see below for comments on boundary asymmetries) and that accelerations are small. These two features cause the initial necking of the bridge (where over pressure initially builds up and the instability develops) to occur roughly at the mid point between the two plates. Generally, wherever the initial necking point forms, the stretching can be considered to be locally symmetric around that point if its location does not accelerate. Then, there is no preferred direction in the reference frame of that point. If that is not the case, asymmetric features in the height profile will have time to develop. The condition for symmetric breakup is thus that the externally imposed acceleration is slower than that induced by surface tension: $\sigma/(\rho h^2 \dot{v})\gg1$, where $h$ is a typical height and $\dot{v}$ a typical acceleration scale related to the movement of the boundaries. In our experiments, while there is always some acceleration of the plates, that acceleration remains small (i.e., since we do not reach high We), and we thus remain in the symmetric regime and do not observe breakup features that are due to acceleration, as explored in~\cite{zhuang_combined_2015}.

For the highest We experiments we observe a systematic asymmetry in the breakup: the pinch-off always first occurs on the left, as seen in Fig.~\ref{fig:low_hi} $(b)$ for a few examples with different We and $L_0$. This asymmetry seems to originate from an asymmetry at the boundaries. Note however that this did not seem to have a significant affect for the earlier breakup dynamics during the universal dynamics, and in particular did not seem to alter the scaling of the satellite bubble size for these We.

\subsection{Realizability of the relevant regime for a liquid bridge}
While satisfying $\text{We}\ll1$ for a liquid bridge requires a smaller set-up and slower speeds than those used in our experiments, it is still possible to keep ${\rm Re}\gg 1$. Indeed, for a water bridge with plate velocities of the order of $v_p\sim 10\, {\rm cm}/{\rm s}$ and boundary radius $h_0=1\, {\rm mm}$ (such that it is below the capillary wavelength), one finds $\text{We}\sim 10^{-1}$ and $\text{Re}\sim 10^2$. Ref.~\cite{zhang_nonlinear_1996} also includes bridges in this regime: using a larger set-up ($h_0\sim 10 {\rm  cm}$) with slower speeds ($v_p\sim 1 {\rm  mm/s}$) such that ${\rm We}\sim 0.001$ and ${\rm Re}\sim 10^2$ (placed in a Plateau tank to achieve a small Bond number).

\bibliography{bubble}

\end{document}